\documentclass[fleqn,usenatbib]{mnras}

\usepackage{newtxtext,newtxmath}
\usepackage{color}

\usepackage[T1]{fontenc}
\usepackage{ae,aecompl}

\usepackage{graphicx}	% Including figure files
\usepackage{amsmath}	% Advanced maths commands
\title[Starfall in "turning on" AGN]{Starfall: A heavy rain of stars in ``turning on'' AGN}

\author[B.McKernan et al.]{B. McKernan$^{1,2,3,4}$\thanks{E-mail:bmckernan at amnh.org (BMcK)} , K.E.S. Ford$^{1,2,3,4}$, M. Cantiello$^{2,5}$, M. Graham$^{6}$, A. S. Jermyn$^{2}$, N.W.C. Leigh$^{7,1}$,
\newauthor T. Ryu$^{8}$ \& D. Stern$^{9}$\\
$^{1}$Department of Astrophysics, American Museum of Natural History, New York, NY 10024, USA\\
$^{2}$Center for Computational Astrophysics, Flatiron Institute, New York, NY 10010, USA\\
$^{3}$Graduate Center, City University of New York, 365 5th Avenue, New York, NY 10016, USA\\
$^{4}$Department of Science, BMCC, City University of New York, New York, NY 10007, USA\\
$^{5}$Department of Astrophysical Sciences, Princeton University, Princeton, NJ 08544, USA\\
$^{6}$California Institute of Technology, 1200 E. California Blvd, Pasadena, CA 91125, USA \\
$^{7}$Departamento de Astronom\'a, Facultad de Ciencias F\'sicas y Matem\'aticas, Universidad de Concepci\'on, Concepci\'on, Chile \\
$^{8}$Max Planck Institute for Astrophysics, Karl-Schwarzschild-Strasse 1, 85748 Garching, Germany\\
$^{9}$Jet Propulsion Laboratory, California Institute of Technology, 4800 Oak Grove Drive, Pasadena, CA 91109, USA \\
}

\date{Accepted XXX. Received YYY; in original form ZZZ}

\pubyear{2021}

\begin{document}
\label{firstpage}
\pagerange{\pageref{firstpage}--\pageref{lastpage}}
\maketitle

\begin{abstract}
As active galactic nuclei (AGN) `turn on', some stars end up embedded in accretion disks around supermassive black holes (SMBHs) on retrograde orbits. Such stars experience strong headwinds, aerodynamic drag, ablation and orbital evolution on short timescales. Loss of orbital angular momentum in the first $\sim 0.1$~Myr of an AGN leads to a heavy rain of stars (`starfall') into the inner disk and onto the SMBH. A large AGN loss cone ($\theta_{\rm AGN,lc}$) can result from binary scatterings in the inner disk  and yield tidal disruption events (TDEs). Signatures of starfall include optical/UV flares that rise in luminosity over time, particularly in the inner disk. If the SMBH mass is $M_{\rm SMBH} \ga10^{8}M_{\odot}$, flares truncate abruptly and the star is swallowed. If $M_{\rm SMBH}<10^{8}M_{\odot}$, and if the infalling orbit lies within $\theta_{\rm AGN,lc}$, the flare is followed by a TDE which can be prograde or retrograde relative to the AGN inner disk.  Retrograde AGN TDEs are over-luminous and short-lived as in-plane ejecta collide with the inner disk and a lower AGN state follows. Prograde AGN TDEs add angular momentum to inner disk gas and so start off looking like regular TDEs but are followed by an AGN high state. Searches for such flare signatures test models of AGN `turn on', SMBH mass, as well as disk properties and the embedded population.
\end{abstract}

%\textcolor{red}{NL: The TDE radius depends on the spin of the SMBH.  It can be up to 10$^9$ M$_{sun}$ above which no TDEs can occur.  At 10$^8$ and less, if the AGN is "on", you will never see these events.  You can cite this paper for this:  https://ui.adsabs.harvard.edu/abs/2018MNRAS.479.3181F/abstract.  Basically, it can only happen secularly, not in a chaotic scenario, so this might be good for your arguments, not completely sure.  Ref is in the refs.bib.}
%BM: Essentially we're ignoring BH spin for the moment. Also, whether you see a TDE flare very much depends on the intrinsic AGN luminosity and the sort of characteristic variability associated with it!


%%%%%%%%%%%%%%%%%%%%%%%%%%%%%%%
\section{Introduction}
Supermassive black holes (SMBHs) in galactic nuclei are surrounded by stars and stellar remnants in nuclear star clusters (NSCs), with millions of solar masses packed within the central few parsecs of
a galaxy.  NSCs, which generally have a spheroidal geometry, are the densest stellar systems known; for a recent review of NSCs, see \citet{Neumayer20}. Active galactic nucleus (AGN) disks are powered by accretion from gas disks onto SMBHs. When AGN disks form (`turn on') around the SMBH, a fraction of the NSC orbits (approximately given by the disk aspect ratio) will be co-planar with the disk. Around half of the initially co-planar embedded orbits should be retrograde compared to the gas angular momentum. 

Stars on prograde orbits in AGN disks experience  different boundary conditions than stars in vacuum. These stars accrete mass and evolve very differently than their cousins in vacuo \citep{MatteoAdam20,MelvynDoug20}. A simple flip of angular momentum orientation leads to a  different stellar fate. Embedded stars on retrograde orbits in AGN disks will experience strong headwinds as they move backwards through the gas, while Bondi accretion is weak due to the large relative velocity of the star with respect to the gas. For retrograde orbiters on near circular orbits, the headwind causes rapid angular momentum loss and a population of low angular momentum stars on retrograde orbits builds up quickly in the inner AGN disk. Under such conditions, mass loss, gas heating, and orbital evolution can occur on short timescales. Among this population in the inner disk, strong dynamical encounters, particularly between binaries and tertiaries \citep[e.g.][]{leigh18, Samsing20}, can quickly fill the AGN loss cone (see \S\ref{sec:AGN_losscone} below) and lead to tidal disruption events (TDEs) in AGN with $M_{\rm SMBH} \la 10^{8}M_{\odot}$ around non-spinning SMBH. 

\begin{figure*}
\begin{center}
\includegraphics[width=0.85\linewidth,angle=0]{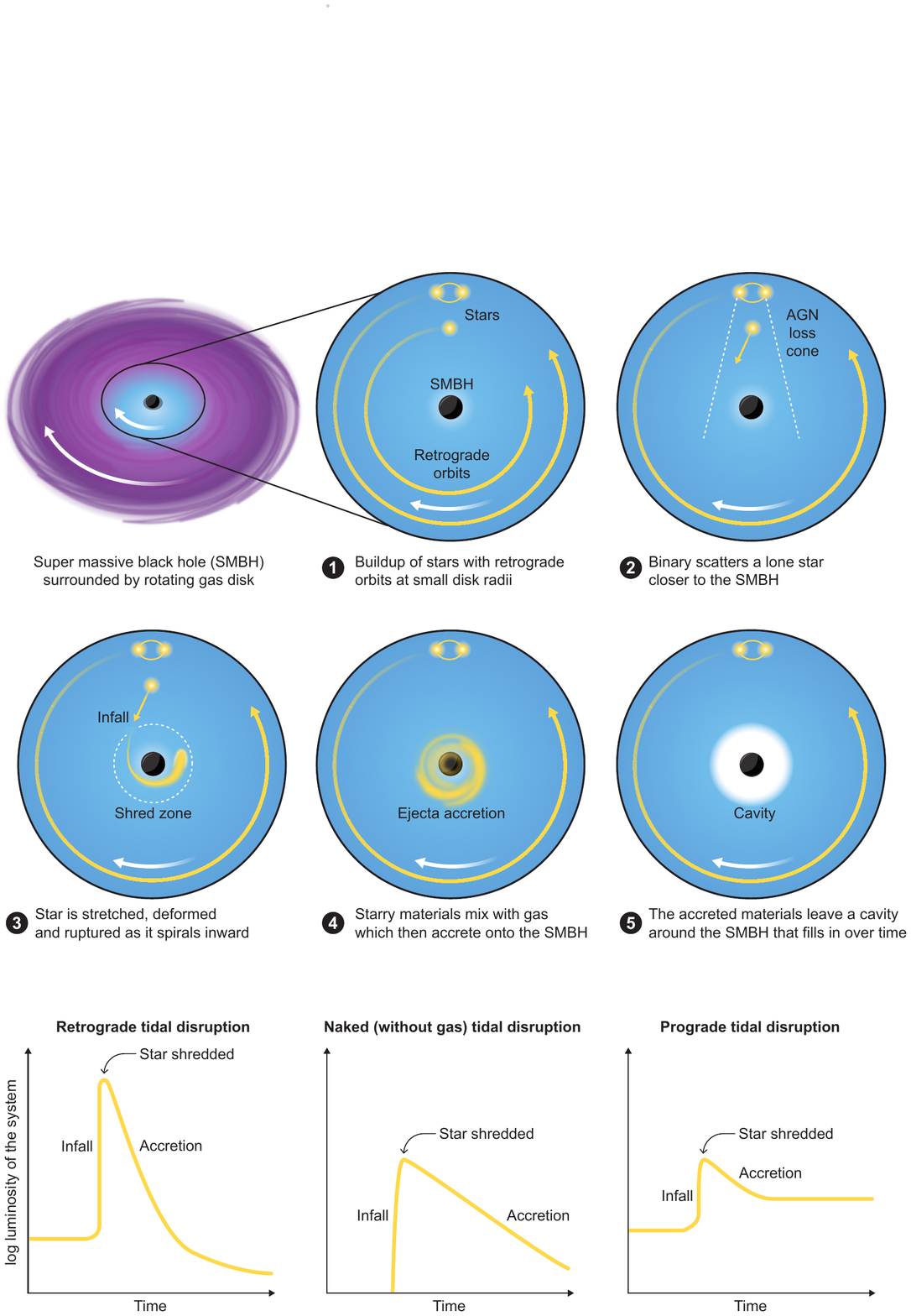}
\end{center}
\caption[BrokenSym]{Cartoon illustrating the model of Starfall. Top left panel shows a gas disk orbiting an SMBH, with bluer, hotter inner regions and redder, cooler outskirts. Panel 1) shows the build up of a low angular-momentum retrograde population of stars at small disk radii. 2) shows a fiducial chaotic scattering from a (2+1) encounter into the AGN loss cone. The scattered star will be disrupted in a prograde or retrograde sense compared to disk gas depending on which side of the SMBH it passes. 3) shows a \emph{retrograde} tidal disruption event in the plane of the disk with tidal debris hitting the inner AGN disk, driving a strong accretion episode in 4), followed by clearing out of a cavity in (5) which will re-fill on the viscous timescale. Bottom row illustrates the qualitative lightcurves we expect from a retrograde TDE (LHS), a 'naked' TDE (middle) and prograde TDE (RHS). A \emph{retrograde} TDE is more luminous than a naked TDE since all of the disrupted stellar mass is bound, together with some of the inner disk mass and fall back occurs over a shorter time than for 'naked' TDEs. Retrograde TDEs are followed by a lower AGN continuum state as the inner disk re-fills on the viscous timescale. Ejecta from a \emph{prograde} TDE adds angular momentum to the inner disk gas, inhibiting inflow while the TDE occurs. Prograde TDEs are around the same luminosity as 'naked' TDEs and are followed by a higher AGN continuum state. (credit: Lucy Redding-Ikkanda/Simons Foundation)
\label{fig:starfall_cartoon}}
\end{figure*}

This paper is structured as follows: We give an overview of the basic model of starfall in \S\ref{sec:overview}. We introduce retrograde orbiters in AGN disks in \S\ref{sec:retro}. We discuss the scouring headwinds and their effects on the star in \S\ref{sec:ablation} and \S\ref{sec:response}. How retrograde orbits evolve due to the scouring headwind is discussed in \S\ref{sec:evolution}, and the AGN disk loss cone due to scattering encounters and/or eccentricity pumping is introduced in \S\ref{sec:AGN_losscone}. Flares due to final star fall through the disk are discussed in \S\ref{sec:starfall}, and TDEs in AGN in the first $\sim 0.1$~Myr after an AGN `turns on' are discussed in \S\ref{sec:tdes}. Some of the consequences of starfall for studies of changing-look AGN and the LIGO AGN channel are discussed in \S\ref{sec:discussion}.

\section{A brief overview of Starfall}
\label{sec:overview}
Fig.~\ref{fig:starfall_cartoon} provides a basic illustration of the model of Starfall and provides a useful guide for the details discussed in this paper. The top left panel shows a gas disk orbiting a SMBH, bluer/hotter in the interior and redder/cooler in the exterior. Panel (1) shows a polar-view of the build up of low angular momentum retrograde orbiters over a relatively short time in the inner disk. Panel (2) shows a polar-view of the result of a chaotic scattering encounter sending a singleton into the AGN loss cone (see \S\ref{sec:AGN_losscone} below). The resulting TDE has approximately equal odds of being prograde or retrograde compared to the inner disk gas, yielding two equally likely possible lightcurves. Panel (3) shows a polar view of a retrograde TDE, including the collision of all ejecta in plane with the inner AGN disk. Panel (4) shows the resulting period of turbulent, enhanced accretion onto the SMBH post-TDE. Panel (5) shows the cavity formed in the inner disk after a retrograde TDE accretion episode, which will fill in on the viscous timescale of the cavity boundary and leads to a lower AGN state post-TDE. The bottom row of panels shows the qualitative difference between expected lightcurves. Left-hand panel bottow row is the lightcurve expected from a retrograde TDE. There is a rise in luminosity from the pre-TDE AGN value during the final infall, followed by a peak corresponding to the TDE event. The accretion episode of a retrograde TDE is more luminous and decays more rapidly than a standard TDE (depicted in middle panel), since fall-back should occur over a much shorter timescale. After a retrograde TDE the AGN continuum state is lower than the pre-TDE state since the inner disk is mostly cleared out and recovers to the pre-TDE state on the viscous timescale set by the cavity inner edge. By contrast the prograde TDE ejecta adds angular momentum to the inner disk ejecta and yields a TDE of comparable luminosity to a 'naked' TDE, followed by a delayed higher AGN continuum state.

\section{Retrograde orbiters in AGN disks}
\label{sec:retro}
How exactly AGN turn on remains unclear. Plausibly, a pulse of low angular momentum gas arrives in the central galactic nucleus, either from a fuel reservoir such as a torus, or from elsewhere in the host galaxy. Signs of possible AGN `turn on' have been observed recently among changing-look or changing-state AGN (CAGN). CAGN exhibit large magnitude changes on short timescales in the optical band \citep[e.g.][]{LaMassa15,Ross18,Graham17} and represent a small fraction of AGN in the local Universe ($z\leq 1$) \citep[e.g.][]{Graham20}. CAGN variability may be due to a large number of possible processes \citep[see e.g.][\& references therein]{Stern18}, but it is possible that some CAGN may be turning on. Other examples of AGN possibly turning-on \citep[e.g.][]{Comerford17} suggest that some AGN are short-lived ($\sim 0.1$Myr), powered by small mass pulses of low angular momentum gas, which are accreted rapidly. If the lifetimes of AGN episodes span $\tau_{\rm AGN}=0.1-100$Myr \citep[see e.g.][]{Schawinski15}, we should expect to see a `turn on' rate of $1/\tau_{\rm AGN}$, or 1 per $10^{5}-10^{8}$ AGN surveyed, per year. As time domain surveys are now photometrically covering $>10^6$ AGN per year, we should expect to see the signature of AGN `turn on' in a small but non-negligible fraction of AGN observed by e.g. ZTF \citep{Graham17} and in the future with LSST. Besides the signatures discussed here, we note that if there was no recent prior AGN episode, newly 'turned on' AGN may not have narrow {[O~{\sc iii]}} lines in their optical spectra.

How AGN disks form is not yet well understood, but we can make some general arguments about dominant effects. A pulse of low angular momentum gas falling from approximately the dust sublimation radius (and site of a possible reservoir/torus of gas) takes the freefall time 
\begin{equation}
    t_{\rm ff} \sim 10^{2}{\rm yr} \left(\frac{R}{0.4{\rm pc}} \right)^{3/2} \left(\frac{M}{10^{8}M_{\odot}}\right)^{-1}
\end{equation}
to arrive at the SMBH, where $t_{\rm ff}$ is also approximately the cloud crushing time \citep[see e.g.][]{Stern18}. On such timescales the cloud will not interact significantly with most orbiters as it drops. The few orbiters that do interact with the cloud should experience a strong impulsive drag force, causing some loss of angular momentum. Once the pulse of low angular momentum gas arrives around the SMBH, we expect the disk should tend to arrange itself from the inside out, since all timescales (dynamical, thermal, viscous, orbital) are shortest at small radii. 

Assuming most NSC orbits are not significantly perturbed until the disk grows outwards to encompass the NSC orbit, we should expect half of the initial population of embedded orbiters will move backwards compared to the AGN disk gas flow, for an isotropic distribution of orbits. That is, on average about half the initial population of embedded objects should have orbital angular momentum anti-parallel to disk gas angular momentum. The geometric fraction of the nuclear cluster that ends up initially embedded in the disk depends on the number of cluster orbits with semi major axes less than or equal to the disk outer radius ($R_{\rm out}$) and with inclination angles less than the average aspect ratio ($H/R$) of the disk. For a large continuous disk extending to the pc-scale torus, with average $H/R \sim 0.05$, this might be $5\%$ of the initial nuclear cluster population, but most disks are likely significantly smaller in radial extent than this.

Likewise, about half of the NSC orbits with semi-major axis less than the disk radius will cross the disk retrograde. Over the AGN disk lifetime, the population of embedded objects within AGN disks grows due to the orbital grind-down and capture of NSC orbiters \citep{1993ApJ...409..592A,Fabj20,MacLeodLin20} as well as in situ star formation \citep{2004ApJ...608..108G,Dittmann2020}. The population of embedded objects in the disk is the result of a competition between population loss due to mergers, scatterings out of the disk, and extreme mass ratio inspirals (EMRIs) onto the SMBH on the one hand, and disk capture, star formation, and stellar evolution on the other. Later additions to the embedded population in the disk should be almost exclusively prograde orbiters (i.e., orbital angular momentum aligned with the disk angular momentum) \citep{MacLeodLin20}.

In principle, the arrival of a globular cluster or dwarf galaxy in an AGN might permit a starfall-like event to occur via the interaction between new spherical orbits and a gas disk. However, most AGN are probably short-lived ($<$ few Myr) \citep[e.g.][]{Schawinski15}, so the likelihood that a random dynamical decay arrival coincides with a short lived AGN episode is small. The likelihood of such a coincidence is larger for the longest lived AGN (the few that might live $\sim 100$~Myr). So it seems much more likely that low angular momentum gas arrives in the nucleus to impact a spherical orbital distribution than the other way round.  Below, we investigate the consequences of an initial population of retrograde stars on the stars themselves and on AGN variability.

\section{Ablation of stellar surfaces by impulsive scouring headwinds}
\label{sec:ablation}
The headwind velocity ($v_{\rm wind}$) faced by a star moving retrograde through the AGN disk on a Keplerian orbit is given by approximately twice the Keplerian orbital velocity, 
\begin{equation}
    v_{\rm wind}(r) \approx 2\left[ GM_{\rm SMBH}\left(\frac{2}{r}-\frac{1}{a}\right)\right]^{1/2}  
\end{equation}
where $r$ is the instantaneous radial distance from the SMBH and $a$ is the orbital semi-major axis.

On a circular retrograde orbit ($e=0$), the star's relative velocity to the disk is supersonic and the headwind creates a bow shock ($R_{\rm shock} \gtrsim R_{\ast}$), where $R_{\rm shock}$ is the bow shock radius and $R_{\ast}$ is the radius of the stellar photosphere. On a more eccentric orbit, the headwind velocity at periastron could be mildly relativistic. In the limit of small $e$, the velocity difference is $\propto e$. But we shall ignore the effect of eccentricity in the discussion that follows. 

The shock lies where the headwind pressure balances either the stellar wind pressure or the stellar atmospheric pressure. The headwind pressure is large so the latter is more likely, giving $R_{\rm shock} \sim R_{\ast}$. The power in the shock is $\mathcal{P}_{\rm shock} \sim R_{\rm shock}^{2} \rho_{\rm wind} v_{\rm wind}^{3}$.
An order-unity fraction of this power goes into energy dissipation by the shock. When that energy is re-radiated. a fraction of order $(R_{\rm shock}/R_{\ast})^{2}$ of it impinges on the stellar surface, adding a luminosity
\begin{align}
    L_{\rm shock} & \approx\pi  R_{\ast}^{2} \rho_{\rm wind} v_{\rm wind}^{3}\\
        &\approx 10^{6} L_\odot \left( \frac{a}{10^{3}r_{g}}\right)^{-3/2}\left( \frac{R_{\ast}}{R_{\odot}}\right)^{2} \left( \frac{\rho_{\rm wind}}{10^{-10}{\rm g\, cm^{-3}}}\right)
\end{align}
to the stellar surface, where $r_{g}=GM_{\rm SMBH}/c^{2} \sim 1\, (M_{\rm SMBH}/10^{8}M_{\odot})\, {\rm AU}$ is the SMBH gravitational radius. The temperature that results at $R_{\ast} \sim R_{\odot}$ is far hotter than for an isolated O-star.  

High gas densities can lead to very efficient cooling in radiative shocks \citep[see e.g.][]{Steinberg18}.  In what follows, we make the simplifying assumption that the re-radiated spectrum from the shock emerges with an overall thermalized form.

\section{Response of the star}
\label{sec:response}
The perturbation or shocking of the surface of stars has been studied in the context of binary companions in the presence of supernovae and pulsar winds. Type~Ia supernova shockfronts hit binary companions with shock velocities $\sim 10^{4}{\rm km/s}$ and less mass ejecta ($\sim 0.5M_{\odot}$) than Type II supernovae \citep[e.g.][]{Wheeler75,Marietta00}, i.e. closer to the impulsive shock conditions that may apply to stars on eccentric retrograde orbits in AGN disks. Stellar companions to pulsars can also be scoured by ablating winds \citep{Phinney88}. Ablation has also been considered in the context of planets migrating onto or colliding with their host stars \citep{JiaSpruit18}. Models of shocks and ablation of stars and planets due to headwinds are a reasonable guide when considering the effects of strong headwinds acting on a star in an AGN disk.

\subsection{Pressure Perturbations}

A star on a retrograde orbit in an AGN disk experiences instantaneous ram pressure $P_{\rm ram} $
\begin{equation}
P_{\rm ram}\approx \rho v_{\rm rel}^{2} \sim 10^{8}\,{\rm dyne\, cm^{-2}}  \left(\frac{\rho_{\rm wind}}{10^{-10}\, {\rm g\, cm^{-3}}}\right) \left(\frac{v_{\rm rel}}{10^{4}\, {\rm km\, s}^{-1}}\right)^{2}
\end{equation}
at its surface, where $v_{\rm rel}$ is the relative velocity between the impinging wind and the star surface. For context, $P_{\rm ram}$ is far lower than the pressure exerted by a supernova shockwave on a nearby binary companion and orders of magnitude smaller than the central pressure in a solar type star ($\sim 2 \times 10^{17}{\rm dyne\, cm}^{-2}$).
As a result, the pressure perturbation to the structure of the star is likely small.

Similarly, we can estimate whether or not the shock pressure-strips the stellar envelope.
Disruption of the stellar envelope can occur when the ram pressure is greater than the binding energy of the envelope \citep{JiaSpruit18} or
\begin{equation}
    v_{\rm wind} > v_{\rm esc} \left(\frac{\rho_{\ast}}{\rho_{\rm AGN}} \right)^{1/2}.
\end{equation}
For solar-type stars, $(\rho_{\ast}/\rho_{\rm AGN})^{1/2} >\rm{O}(10^{5})$ typically, where $\rho_{\ast}$ is the average stellar density, so the envelope will never be stripped.

However, for the envelopes of evolved stars
\begin{eqnarray}
    v_{\rm wind} > 3000\, {\rm km\, s}^{-1} \left(\frac{v_{\rm esc}}{100\, {\rm km\, s}^{-1}}\right) \left( \frac{\rho_{\ast}}{10^{-7}\, {\rm g\, cm}^{-3}}\right)^{1/2} \nonumber \\
    \times \left(\frac{\rho_{\rm AGN}}{10^{-10}\, {\rm g\, cm^{-3}}} \right)^{-1/2}
\end{eqnarray}
so loosely bound stellar envelopes are very quickly stripped away on most retrograde orbits in AGN disks, enriching disk metallicity.

\subsection{Temperature Perturbations}

Much more significant is the luminosity of the shock, which can easily be orders of magnitude greater than the intrinsic luminosity of the star.
Because the shock is radiative it produces a thermal spectrum at temperature 
\begin{align}
    T_{\rm s} &\approx \left(\frac{L_{\rm shock}}{4\pi R_{\rm shock}^2 \sigma}\right)^{1/4}\approx 2\times 10^5\,\mathrm{K} \left( \frac{a}{10^{3}\, r_{g}}\right)^{-3/8}\left( \frac{\rho_{\rm wind}}{10^{-10}\, {\rm g\, cm^{-3}}}\right)^{1/4},
\end{align}
which is just the effective temperature of a black body of radius $R_{\rm shock} \sim R_\ast$ carrying the shock luminosity.

This radiation imposes a new and much hotter effective boundary condition on the shock-facing side of the star.
If this boundary condition applied everywhere and not just beneath the shock, the star would form a near-isothermal layer running down to approximately the point where the unperturbed star would have $T \approx T_{\rm s}$~\citep{2006ApJ...650..394A,2000A&A...360..969R}.
This reduces the density of the material in that layer by increasing its temperature, producing a change in the radius of the star given by
\begin{align}
    \frac{4}{3} \pi \left(R_{\ast,\rm f}^3 - R_{\ast,\rm i}^3\right) = \int \frac{1}{\rho_{\rm f}} - \frac{1}{\rho_{\rm i}} dm,
\end{align}
where $\rho$ is the density of material in the star, $m$ is the stellar mass coordinate, and the subscripts $i/f$ denote the initial and final state respectively.
Assuming the radius does not change too much, the left-hand side may be approximated by $4\pi R_\ast^2 \Delta R_\ast$, and the right-hand side by $-\int \Delta \ln \rho dm/\rho$.
This may be further approximated by
\begin{align}
   \Delta R_\ast \approx -\delta R_{\rm iso} (\Delta \ln \rho)_{\rm iso} \approx \delta R_{\rm iso} (\Delta \ln T)_{\rm iso},
\end{align}
where $\delta R_{\rm iso}$ is the unperturbed thickness of the isothermal layer and $(\delta \ln T)_{\rm iso}$ is the mass-averaged change in $\ln T$ in that layer.
The latter is order-unity because $T_{\rm s}$ is much greater than the unperturbed $T_{\rm eff}$ of the star and we only average over the layer in which $T_{\rm s} > T_{\rm unperturbed}$.
Hence
\begin{align}
   \Delta R_\ast \approx \delta R_{\rm iso}.
\end{align}
For a solar-type star and $T_{\rm s} \approx 2\times 10^5\,\mathrm{K}$ this corresponds to a change of roughly $0.03 R_\odot$, which is small enough that we can neglect it.

Furthermore the boundary condition only applies on the shock-facing side of the star, and so that side is preferentially heated.
This heating generates large temperature differences along isobars, which in turn drive circulation currents that act to redistribute a portion of the incident radiation.
This is a complex multidimensional process, but order-of-magnitude considerations suggest that the column density at which the temperature becomes spherically symmetric is of order~\citep{2017MNRAS.469.1768J}
\begin{align}
    \Sigma_{i} \approx \frac{G M_\ast \sigma T_{\rm s}^4}{c_s^5} \approx 10^{9}\, \mathrm{g\,cm^{-2}} \left(\frac{T_{\rm s}}{2\times 10^5\,\mathrm{K}}\right)^{3/2}\left(\frac{M_\ast}{M_\odot}\right),
    \label{eq:sig}
\end{align}
where $c_s$ is the sound speed of stellar material at $T_{\rm s}$ and where we have assumed a composition of fully-ionized Hydrogen.

The column density in equation~\eqref{eq:sig} is of order $10^{-2}\, M_\odot/R_\odot^2$ and occurs deeper down than the point where $T(\Sigma)=T_{\rm s}$.
As a result, it is a good approximation to say that the star never redistributes a substantial fraction of the incident heat, and hence the ``night'' side remains much cooler than the shock side.
We may therefore neglect perturbations to the ``night'' side of the star, and limit our treatment of heating and expansion to the shock-facing side.

\subsection{Thermal Ablation}

The shock luminosity also causes mass loss.
Following~\citet{2021MNRAS.500.1592G}, we find that the ratio of the characteristic particle temperature $T_{\rm ch}$ to the virial temperature $T_{\rm g}$ is
\begin{align}
    \frac{T_{\rm ch}}{T_{\rm g}} \approx 2\left(\frac{T_{\rm s}}{2\times 10^5\,\mathrm{K}}\right)^{10/3}\left(\frac{R_\ast}{R_\odot}\right)^{5/3}\left(\frac{M_\ast}{M_\odot}\right)^{-1},
\end{align}
which puts the system on the border between their intermediate and cold wind scenarios.
The resulting mass flux is given by equation (10) of~\citet{2021MNRAS.500.1592G} as
\begin{align}
    \dot{M} \approx 4\pi R_\ast^2 \frac{\sigma T_{\rm s}^4}{c_s v_{\rm g}} \approx 0.3\, M_\odot\, \mathrm{yr}^{-1},
\end{align}
where $v_{\rm g} \equiv \sqrt{G M_\ast/R_\ast}$ is the virial velocity, we have assumed a shock temperature of $2\times 10^5\,\mathrm{K}$, and we have assumed that the shock encompasses an order-unity fraction of the solid angle as seen from the center of the star.

We note that the energy dissipation due to the shock, per orbit, is a tiny fraction of the total orbital energy of the star. For a $1M_{\odot}$ star moving for a time $t$, $L_{\rm shock} t / E_{\rm orb}\simeq 10^{-4}(\rho_{\rm wind}/10^{-10}{\rm g cm^{-3}})(\bar{\rho}_{\ast}/1{\rm g\, cm^{-3}})(a/1000r_{\rm g})(R_{\ast}/R_{\odot})^{-1}(t/P)$ where $E_{\rm orb}=GM_{\rm SMBH}M_{\star}/2a$, $\bar{\rho_{\ast}}$ is the average stellar mass and $P$ is the orbital period of the star.

\section{Evolution of retrograde orbits}
\label{sec:evolution}
The headwind experienced by retrograde orbiters imparts momentum $-{\rm M}_{\rm wind} {\rm v}_{\rm wind}$ to the star. For retrograde stars on highly eccentric orbits, if density in the inner disk is approximately constant \citep[e.g.][]{Sirko03}, the momentum imparted is greatest at periapse. As a result, eccentric orbits in relatively constant density inner disks will tend to circularize at smaller radii in a short time. This geometric drag on objects with material surfaces (i.e. stars) acts against eccentricity pumping that can occur for objects on retrograde orbits where the dominant drag is Bondi-Hoyle-Lyttleton \citep[see e.g.][]{Secunda20b}. The loss of angular momentum ($L=M_{\ast}{\rm v}_{\ast}{\rm a}_{\ast}$) by a star on a quasi-circular retrograde orbit is
\begin{equation}
    \frac{dL}{dt}=-\frac{dM_{\ast}}{dt}{\rm v}_{\ast}{\rm a}_{\ast} -M_{\ast}\frac{d{\rm v}_{\ast}}{dt}{\rm a}_{\ast} -M_{\ast}{\rm v}_{\ast}\frac{d{\rm a}_{\ast}}{dt}
\end{equation}
where $-dM_{\ast}/dt$ is the mass lost by ablation from the star, $-d{\rm v}_{\ast}/dt$ is the deceleration of the star due to the headwind and $-d{\rm a}_{\ast}/dt$ is the change in orbital semi-major axis for the star. 

From above, $-dM_{\ast}/M_{\ast}$ is expected to be very small per orbit, and assuming the semi-major axis change per orbit ($d{\rm a}_{\ast}/dt$) is small,  the angular momentum of the star will drop by $\sim 2M_{\phi}{\rm v}({\rm a}_{\ast}) {\rm a}_{\ast}$ per orbit, where  $M_{\phi}$ is the mass of (anti-) co-rotating gas in an annulus of cross-section $\pi R_{\ast}^{2}$ located at semi-major axis ${\rm a}$ from the SMBH. Since 
\begin{equation}
    M_{\phi} \sim 10^{-4}M_{\odot}\left( \frac{R_{\ast}}{R_{\odot}}\right)^{2} \left(\frac{{\rm a}}{10^{3}r_{g}} \right)\left( \frac{M_{\rm SMBH}}{10^{8}M_{\odot}}\right) \left( \frac{\rho}{10^{-10}\, {\rm g\, cm^{-3}}}\right)
\label{eq:mphi}
\end{equation}
then after $\sim 10^{4}$ orbits ($\sim 0.03$\, Myr in this example), the angular momentum of the retrograde orbiter is cancelled and it plunges into the SMBH, assuming gas turbulence and accretion inflow are sufficient to refill the anti-co-rotating gas volume. More precisely, as the retrograde star loses sufficient angular momentum, its orbital velocity decreases and its semi-major axis drops ($d {\rm a}_{\ast}/dt <0$). AGN density tends to drop in the hot innermost disk, so $M_{\phi}$ encountered by the star decreases, but the headwind magnitude ($v_{\rm rel}$) increases. If $(M_{\phi},\rho)$ drop sufficiently sharply with $d {\rm a}_{\ast}/dt$, a population of low angular momentum retrograde stars could accumulate at small disk radii (panel (1) in Fig.~\ref{fig:starfall_cartoon}). This population can generate localized disk heating and will impact the expected population of extreme mass ratio inspirals (EMRIs).

EMRIs occur when a stellar mass BH accretes onto an SMBH via GW emission. Importantly, if EMRIs are to be detectable by LISA, they must be long, uninterrupted inspiral mergers \citep[e.g.][]{Pau18}. A large population of stars on small semi-major axis orbits near the SMBH would cause sufficient orbital perturbations to disrupt any detectable EMRIs, at least until the stellar population is accreted onto the SMBH via GW decay or tidal disruption. 

Some retrograde stars are likely to end up quickly losing most of their angular momentum and plunging inwards to the SMBH. While Sun-like stars at $\sim 10^{3}r_{g}$ decay in $10^{4}$ orbits in the example above, A-stars decay in half the time and O-stars decay in about $1/3$ the time. So, for a given orbital semi-major axis, we expect MS stars with the largest radii to decay onto the SMBH first, followed by smaller radius (less massive) stars. 

How exactly the orbit decays changes both the fate of the star, and the appearance of the AGN. If the star is not directly scattered into the AGN disk loss cone (see below) it could experience Roche-lobe overflow (RLOF) onto the SMBH in the innermost disk. Such a state might yield an enhanced accretion episode onto the SMBH, with an associated color change as the inner disk temperature ($\propto \dot{M}_{\rm SMBH}^{1/4}$) and luminosity ($\propto \dot{M}_{\rm SMBH}$) increase. These state changes might be used to count the typical number of embedded stellar objects in AGN disks ($N_{\star}$) and the typical orbital decay time ($t_{\rm decay}$) since the order-of-magnitude rate of RLOF events could be as much as $\sim {\rm O}(10^{3}){\rm yr^{-1}} (N_{\rm AGN}/10^{6})(\overline{N}_{\star}/10^{2})(\overline{t}_{\rm decay}/0.1{\rm Myr})$. Below, we are interested in the fate of dynamically interacting retrograde stars and the transient signatures they may generate.

\section{The AGN disk loss cone: (2+1) encounters}
\label{sec:AGN_losscone}
The basic mechanism within AGN disks that can lead to TDEs is scatterings from chaotic dynamical encounters at relatively small disk radii into an AGN loss cone (panel (2) in Fig.~\ref{fig:starfall_cartoon}). Strong dynamical encounters within AGN disks lead to a chaotic range of outcomes, including scattering \citep[e.g.][]{Samsing20, Secunda20a}. The largest range of changes in direction of an orbiter comes from chaotic 3-body encounters due to, e.g., a binary closely encountering a tertiary object \citep[e.g.][]{StoneLeigh19}. So, if a population of retrograde orbiters grows in the inner disk, binaries can form once their mutual binding energy at encounter is greater than their individual kinetic energies. Once this happens, there is a loss cone for strong scattering encounters which can be parameterized as \citep[see e.g.][]{McK21a}
\begin{eqnarray}
    \theta_{\rm AGN,lc} &\sim& 3 \times 10^{-2} \left( \frac{H/a}{0.01}\right)^{-1}\left(\frac {M_{\rm SMBH}}{10^{8}M_{\odot}}\right)^{-2/3} \nonumber \\
    &\times& \left(\frac {M_{\ast}}{M_{\odot}}\right)^{2/3}\left(\frac {R_{\ast}}{R_{\odot}}\right)^{2}\left(\frac {a}{100r_{g}}\right)^{-2}.
    \label{eq:theta2}
\end{eqnarray}
Since $\theta_{\rm AGN, lc} \propto a^{-2}$, scatterings in the inner AGN disk where $a$ is small have the largest AGN loss-cone. So, if a population of retrograde orbiters grows quickly in the inner AGN disk due to orbital decay, the possibility of scatterings increases and the loss cone ($\theta_{\rm AGN,lc}$) becomes large, leading to a potentially high rate of TDEs.  
%\textcolor{red}{NL: 
We expect the singles entering the AGN loss cone as a result of scattering to be divided evenly between prograde and retrograde plunging orbits (i.e. scattering to the either the left or right of the SMBH as seen from the scatterer), yielding prograde or retrograde TDEs (e.g. panel (3) in Fig.~\ref{fig:starfall_cartoon} depicts a retrograde TDE). 

\section{Starfall flares in early AGN}
\label{sec:starfall}
In the last stages of a star plunging into the innermost disk, we can write the freefall time as 
\begin{equation}
    t_{\rm ff} \approx 8\, {\rm mo} \left( \frac{M_{\rm SMBH}}{10^{8}M_{\odot}}\right) \left( \frac{a_{0}}{10^{3}r_{g}}\right)^{3/2}
\end{equation}
where $a_{0}$ is the initial orbital semi-major axis from which plunge begins. During infall, heating of the disk due to the shock produced by the infalling star causes the shock luminosity to rise to 
\begin{eqnarray}
   L_{\rm shock} &\approx& 10^{42}\, {\rm erg\, s}^{-1} \left( \frac{a_{0}}{10^{3}r_{g}}\right)\left( \frac{M_{\rm SMBH}}{10^{8}M_{\odot}}\right)\left( \frac{R_{\ast}}{R_{\odot}}\right)^{2} \nonumber \\
   &\times& \left( \frac{v_{\rm rel}}{2 \times 10^{5}\, {\rm km\, s}^{-1}}\right)^{2} \left( \frac{\rho_{\rm wind}}{10^{-10}\, {\rm g\, cm^{-3}}}\right).
\end{eqnarray}
Starfall therefore generates a long, slow-rise flare to peak luminosity $O(10^{42}\, {\rm erg\, s}^{-1}\, (R_{\ast}/R_{\odot})^{2}$) after $O(t_{\rm ff})$. Once in the inner disk, if the star encounters other decayed retrograde orbiters, scattering with a large loss cone ($\theta_{\rm AGN,lc}$) can occur, followed by either: a TDE (if $M_{\rm SMBH}\la 10^{8}\, M_{\odot}$) a sudden swallowing of the star by the SMBH (if $M_{\rm SMBH} \ga 10^{8}\, M_{\odot}$), or escape of the scatterer to large distances. In the later case, a return to periastron at high eccentricity may be sufficient to initiate a TDE on return to the scattering point.

Starfall flares are most detectable for O-type stars falling through dense AGN disks, for which a flare lasting $t_{\rm ff}$ rises to a peak luminosity $\sim 10^{44}\, {\rm erg\, s}^{-1}$, which could be detectable in large-scale AGN surveys \citep[e.g.][]{Graham17, Graham20}.

How starfall flares emerge at the AGN photosphere depends on $R_{\ast}$, the disk aspect ratio ($H/a$) and the optical depth of the disk ($\tau$). For similar considerations regarding electromagnetic counterparts to shock breakouts from AGN disks, see the discussion in  \citet{Perna21}. The diffusion time for an electromagnetic signal to emerge from a dense, optically-thick AGN disk mid-plane is
\begin{equation}
    t_{\rm diff} \sim 2\, {\rm yr} \left( \frac{\tau}{10^{4}}\right) \left( \frac{H/a}{0.01}\right)\left( \frac{a}{10^{3}r_{g}}\right) \left( \frac{M_{\rm SMBH}}{10^{8}M_{\odot}}\right)
\end{equation}
so the early stages of the plunge will not be detectable. However, the last $O(10^{2}$)\, $r_{g}$ of freefall should take $O$(week) around a $\sim 10^{8}M_{\odot}$ SMBH if there is a total loss of angular momentum. For fiducial disk models \citep[e.g.][]{Sirko03}, $H/a$ in the innermost  disk increases to $\sim 0.1$, but ($\tau, a_{\ast}$) decrease by orders of magnitude from their starting values. As the star approaches the edge of the inner disk at $\sim 6r_{g}$, $t_{\rm diff}$ can drop to days. So the final stages of starfall can emerge as a short timescale ($\sim $week) rising flare at approximately $L_{\rm shock}$. What happens once the plunging star arrives at the SMBH depends on $M_{\rm SMBH}$ and whether the velocity vector of the star lies within $\theta_{\rm AGN,lc}$.

\section{A hard rain of bright TDEs in $<0.1$Myr}
\label{sec:tdes}
Stars can be tidally disrupted by non-spinning SMBHs with $M_{\rm SMBH} \la 10^{8}M_{\odot}$ \citep{Rees88}. For TDEs occuring in vacuo, approximately half of the disrupted stellar mass is expected to escape on hyperbolic orbits at $v_{\infty} \sim 10^{4}\, {\rm km\, s}^{-1}(M_{\ast}/M_{\odot})^{1/2}(R_{\ast}/R_{\odot})^{-1/2}(M_{\rm SMBH}/10^{6}M_{\odot})^{1/6}$ \citep{Rees88}. Around half the disrupted stellar mass remains bound and fall back occurs over $dM/dt \propto t^{-5/3}$ \citep{Phinney89}. Two bound streams collide near apocenter at $r_{a} \sim R_{\ast} q^{-2/3}$ on a characteristic timescale $t_{\rm TDE} \sim (r_{a}^{3}/2GM_{\rm BH})^{1/2}$ \citep[e.g.][]{Piran15} although rapid circularization can also occur \citep[e.g.][]{Rees88,EvansKochanek89,Guillochon15}. 

However, the situation is quite different for stars plunging onto an SMBH through an AGN disk (panel (2) in Fig.~\ref{fig:starfall_cartoon}). Ignoring SMBH spin, if a star falls onto a $\ga 10^{8}M_{\odot}$ SMBH, it will simply be swallowed; so in what follows, we assume $M_{\rm SMBH} \la 10^{8}M_{\odot}$. Note that for $M_{\rm SMBH}\gtrsim 5\times10^{6}M_{\cdot}$ the rate at which stars are swallowed whole (direct capture) is higher than for TDEs \citep{Taeho20a,Taeho20b}. When a star is scattered into the AGN loss cone, the result can be (approximately equally) prograde or retrograde TDEs, which we discuss below. 

\subsection{Retrograde TDEs}
In the case of a retrograde TDE, ejecta ploughs directly into the inner disk, rapidly  cancelling angular momentum and driving a short-lived, luminous pulse of accretion onto the SMBH (panels (3) and (4) in Fig.~\ref{fig:starfall_cartoon}). Broadly the lightcurve we expect is a shorter, sharper (more luminous) TDE than in naked galactic nuclei, followed by a low AGN state (bottom row, left panel in Fig.~\ref{fig:starfall_cartoon}). 

The fastest, otherwise unbound, debris loses kinetic energy from initial values of $\sim 0.5 M_{\ast} v_{\infty}^{2}$ and exchanges angular momentum with inner disk gas. Ejecta travel an average distance 
\begin{equation}
    R_{d}(r_{g}) \sim \left(\frac{M_{\rm d}}{M_{\ast}}\right)\left(\frac{c}{v_{\infty}} \right)^{2}
\end{equation}
into the disk after sweeping up mass comparable to ejecta mass, where $M_{\rm d}$ is the mass of gas in the inner disk out to radius $R_{d}$, and can be parameterized as
\begin{equation}
    M_{\rm d} \approx  0.03M_{\odot} \left(\frac{M_{\rm SMBH}}{10^{7}M_{\odot}}\right)\left(\frac{R_{\rm d}}{10^{2}r_{g}}\right)^{2}\left(\frac{H/r}{0.05}\right)\left(\frac{\rho}{10^{-10}{\rm g\, cm^{-3}}}\right).
\end{equation}
Thus, the characteristic TDE timescale is now $t_{\rm TDE,AGN} \sim (R_{d}^{3}/2GM_{\rm BH})^{1/2}$ where $R_{d} \ll r_{a}\sim R_{\ast}q^{-2/3}$. The characteristic timescale of AGN TDE decay compared to naked TDE decay is
\begin{eqnarray}
    \frac{t_{\rm TDE,AGN}}{t_{\rm TDE}} &\sim & \left(\frac{M_{d}}{M_{\ast}}\right)^{3/2}\left(\frac{c}{v_{\infty}}\right)^{3} q \left(\frac{r_{g}}{R_{\ast}}\right)^{3/2} \nonumber \\
    &\sim& 10^{-3} \left(\frac{M_{d}/M_{\ast}}{0.03}\right)^{3/2}\left(\frac{c/v_{\infty}}{30}\right)^{3}\left(\frac{q}{10^{-7}}\right)\left(\frac{r_{g}/R_{\ast}}{20}\right)^{3/2}
\end{eqnarray}
and so AGN TDE decay is $O$(days) rather than $O$(yrs). Unbound ejecta take time ($t_{d}$) 
\begin{equation}
t_{\rm d} \sim \frac{R_{d}}{v_{\infty}} \sim 0.01{\rm yr}\left(\frac{M_{\rm SMBH}}{10^{7}M_{\odot}}\right)\left(\frac{R_{d}}{10^{2}r_{g}}\right)\left(\frac{v_{\infty}}{10^{4}{\rm km\, s}^{-1}}\right) 
\end{equation}
to cross the inner disk. Since $t_{\rm d} \sim T_{\rm orb}(10^{2}r_{g})$, the ejecta encounters approximately all the mass $M_{\rm d}$ for this parameterization. 

Inner disk gas $<R_{d}$ loses angular momentum $\sim 0.5M_{\ast} v_{\infty} \overline{a}$ where $\overline{a}\sim R_{d}/2$ is the average disk radius that the ejecta reach. Therefore a mass of gas $\sim (M_{\rm d}/4) (M_{\ast}/M_{\rm d})(v_{\infty}/v_{\rm orb}(\overline{a}))$ is delivered onto the SMBH in time $t_{d}$ yielding an accretion rate

\begin{eqnarray}
    \dot{M}_{\rm SMBH} &\sim& \frac{M_{\ast}}{4t_{d}}\left( \frac{v_{\infty}}{v_{\rm orb}}\right) \nonumber \\
    &\sim & 0.25\, M_{\odot}{\rm mo}^{-1} \left( \frac{M_{\rm SMBH}}{10^{7}M_{\odot}}\right)\left( \frac{M_{\ast}}{M_{\odot}}\right)\left(\frac{v_{\infty}/v_{\rm orb}}{0.1}\right)
\end{eqnarray}
which is $\times 10$ Eddington ($\dot{M}_{\rm Edd}$) for this parameterization. The associated viscous timescale over which this mass can be accreted is \citep{Stern18}
\begin{equation}
    t_{\nu} \sim 6\, {\rm mo} \left(\frac{h/R}{0.1}\right)^{-2}\left( \frac{\alpha}{0.05}\right)^{-1} \left(\frac{M_{\rm SMBH}}{10^{7}M_{\odot}}\right)\left(\frac{a}{50r_{g}}\right)^{3/2}
\end{equation}  
leading to a draining of the inner disk on this timescale, followed by a lower accretion state as the inner accretion disk recovers (see retrograde lightcurve sketch in the leftmost panel of the bottom row in  Fig.~\ref{fig:starfall_cartoon}). The actual  behaviour of the unbound and bound gas is likely to be more complicated than the simple model presented here and will require detailed simulations of pro- and retrograde TDEs impacting Keplerian flows.

The maximum temperature of the innermost stable circular orbit (ISCO) is given by \citep[e.g.][]{Zimmerman05}
\begin{equation}
T_{\rm ISCO} =f \left( \frac{3GM_{\rm SMBH}\dot{M}}{8\pi r_{\rm ISCO}^{3}\sigma}\right)^{1/4}
\end{equation}
where $r_{\rm ISCO}$ is the innermost stable circular orbit (ISCO) of the accretion disk and $f$ is a parameter $O$(1) that modifies the spectrum from blackbody. $T_{\rm ISCO}$ can be parameterized as:
\begin{eqnarray}
T_{\rm ISCO} &\approx& 5 \times 10^{5}K
\left(\frac{M_{\rm SMBH}}{10^{7}M_{\odot}}\right)^{-1/4} \left(\frac{\dot{M}}{0.1\dot{M}_{\rm Edd}}
\right)^{1/4} \left(\frac{\eta}{0.1}\right)^{-1/4} \nonumber\\ &\times& \left(\frac{r_{\rm ISCO}}{6 r_{g}}\right)^{-3/4} \left(\frac{f}{2}\right)
\end{eqnarray}
where $\dot{M}$ is a fraction of the Eddington accretion rate, $\eta \sim 0.1$ is the accretion efficiency and $\dot{M}_{\rm SMBH} \sim 0.03M_{\odot}\, \rm{yr}^{-1}$ here, i.e. most of the inner disk mass $M_{\rm d}$ is accreted in a year for our choice of parameters.

\subsection{Prograde TDEs}
Prograde TDEs result from the star entering the AGN loss-cone and slingshotting around the SMBH on a path that matches the flow of the inner disk gas. As a result, TDE ejecta \emph{adds} angular momentum to gas in the inner disk, temporarily pushing the inner disk outwards as the bound part  of the stellar mass is consumed. Thus, we expect a qualitatively different lightcurve for prograde TDEs compared to retrograde TDEs (rightmost panel of the bottom row in Fig.~\ref{fig:starfall_cartoon}). 

Unbound prograde TDE ejecta adds angular momentum and mass to inner disk gas while bound TDE ejecta falls back and is consumed. As a result we expect a TDE-event that initially looks more like a regular 'naked' TDE, but is followed by a higher AGN state than pre-TDE as the extra mass of gas in the inner disk ($\sim M_{d}+0.5M_{\ast}$) accretes on the viscous timescale ($t_{\nu}$) of the inner disk. The accretion rate post-prograde TDE is then O($(M_{d}+0.5 M_{\ast})/t_{\nu}$. For $t_{\nu}/t_{d} \sim O(10^{2})$ as parameterized above, the post-TDE accretion rate is  O(year) after the prograde TDE at a higher rate than before the TDE, leading to a higher AGN state post-prograde TDE.

\subsection{AGN-TDE colours and metallicities}
 If TDE debris is restricted to the innermost disk, the luminosity increase should appear mostly in the UV and blue optical bands. Assuming a default thin-disk temperature profile  $T\propto r^{-3/4}$, this profile will steepen in the inner disk, with a break around the radius where the ejecta mass interacts with the interior disk mass.  If the star envelope has experienced substantial stripping during starfall, the resulting TDE may contain less H relative to He or other metals \citep[see e.g.][]{Arcavi14,Hung17}.
 
Non-AGN TDE source-frame optical/UV luminosities are typically $L_{\rm o/u} \sim {\rm O}(10^{43}){\rm erg/s}$ \citep{Sjoert18}, with estimated bolometric luminosities $L_{\rm bol} \sim {\rm O}(10^{45}){\rm erg/s}$ and decay over O(yr) timescales \citep{Gezari08}, for total energy O($10^{51-52}{\rm erg}$). We expect prograde AGN TDEs to have comparable luminosities and decay rates, followed by a higher AGN state post-TDE as the inner disk, including ejecta mass, loses angular momentum and accretes on the viscous timescale. Retrograde AGN TDEs should have significantly higher luminosities than regular or prograde TDEs, this is because a similar magnitude of energy is released over a much shorter timescale. Thus, retrograde TDEs should have optical/UV luminosities of $L \sim {\rm O}(10^{44}){\rm erg/s}$ or higher, with rapid (months) decay to a lower AGN continuum state as the inner disk refills on the viscous timescale.
 
\subsection{AGN-TDE rate}
The TDE rate for quiescent galactic nuclei is expected to be $O(10^{-4})\, {\rm yr^{-1}}$ \citep{StoneMetzger16}. In AGN that are turning on around SMBH with masses $M_{\rm SMBH}<10^{8}M_{\odot}$, the rate of TDEs is a function of the median disk gas density and the number of retrograde stars embedded in the disk. To order of magnitude, assuming $N_{\star} \sim O(100)$ stars embedded in the initial disk, around $N_{\star}/2 \sim O(50)$ of these stars will have retrograde orbits that decay in $\tau_{\rm decay} <0.1$Myr. Assuming this population interacts and scatters at small disk radii, we should expect an approximate TDE rate of ${\cal{R}}_{\rm TDE} \sim 5 \times 10^{-4}{\rm AGN}^{-1}{\rm {yr}^{-1}} (N_{\star}/10^{2})(\tau_{\rm decay}/0.1{\rm Myr})$ where $\tau_{\rm decay}$ is the typical orbital decay time. Therefore in a large sample of $O(10^{6})$ AGN we should expect $O(500) {\rm yr}^{-1} (N_{\star}/10^{2})(\tau_{\rm decay}/0.1{\rm Myr})$ from turning on AGN. Since scatterings are chaotic, we expect  approximately half of these AGN TDEs should be prograde (which look more like 'naked' TDEs) and half should be retrograde (significantly more luminous and with faster decay).  Thus, a search for TDE-like events in AGN lightcurves can help constrain the AGN disk density (via $\tau_{\rm decay}$) and $N_{\star}$, the number of embedded objects in AGN disks. 

\section{Discussion}
\label{sec:discussion}
How AGN `turn on' is still unknown. For our purposes, we assume that quiescent galactic nuclei do not possess a significant gas disk at ${\rm a}<10^4 r_g$, and an AGN `turns on' when a sufficiently large mass reservoir arrives in that region and forms an accretion (thus, luminous) disk. Stars on retrograde orbits embedded in AGN disks that are in the process of forming, or `turning on', have several important consequences, for the stars, the disk and for observations. 

First, retrograde stars experience a strong headwind that can become mildly relativistic at high eccentricity. The heating of the stellar surface causes the star to lose mass continuously as it orbits. Over time, momentum loss due to the headwind causes orbits to circularize and decay over $<0.1$Myr, leading to a population build-up in the inner disk of retrograde orbiters. Strong dynamical interactions (particularly between a binary and an additional orbiter) among this population can fill the AGN loss cone and produce a very high rate of luminous, short-lived TDEs in AGN in their first $<0.1$Myr. 

As the AGN persists after the first $0.1$Myr starfall, we should expect additional stellar scatterings into the AGN loss cone ($\theta_{\rm AGN,lc}$) from interactions between embedded prograde orbiters. The population $N_{\star}/2$ of prograde embedded stars can grow, migrate and interact. Additional stars can be captured by the AGN disk particularly for large, long-lived, dense disks. If for a given AGN we expect $O(N_{\star}/2)$ TDEs in the first $0.1$Myr and possibly up to $O(N_{\star}/2$ TDEs in the remaining $\tau_{\rm AGN}$. Thus, the distribution of TDE events in AGN can allow us to test the typical median lifetime of AGN. In this parameterized example, around half of the AGN TDEs occur in the first 0.1Myr and the remainder occur in the rest of the AGN lifetime ($\tau_{\rm AGN}$). Thus, the expected rate of starfall TDEs is $O(5 \times 10^{-4(-6)}(N_{\star}/10^{2}){\rm yr}^{-1}{\rm AGN}^{-1})$ for $\tau_{\rm AGN} \sim 1(10)$Myr, where we have ignored  any additional population of stars captured by the disk during $\tau_{\rm AGN}$. 

Approximately half of the TDEs that result from (2+1) scatterings into the AGN loss cone will be retrograde, i.e. their ejecta has negative angular momentum compared to the inner disk gas, and half the TDEs will be prograde. This leads to a difference in the expected observed signature. On one hand, the unbound ejecta from retrograde TDEs will cancel angular momentum of the inner disk and drive a super-luminous TDE event over a short timescale, followed by a lower AGN continuum state. On the other hand, the unbound ejecta from prograde TDEs will add angular momentum to the inner disk and tend to drive the innermost portions of the disk outward, leading to a more 'typical' early TDE profile, followed by a high AGN state, possibly similar to the recent nuclear transient event ASSASN-17jz \citep{Holoien21}.

Changing-look or changing-state AGN (CLAGN) are a new category of AGN exhibiting large magnitude changes on short timescales in the optical band \citep[e.g.][]{LaMassa15,Ross18,Graham17}. Searching for evidence for TDE-like events in CAGN lightcurves is therefore a useful test of whether any CAGN have recently `turned on'. Radio emission from the interaction between unbound ejecta in TDEs and disk gas can be generated and may be detectable. We leave a detailed analysis of this emission to future work. An additional indicator of extreme `youth' may also include missing narrow {[O~\sc{iii}]} lines in the optical AGN spectrum, since the light travel time to the Narrow Line Region (NLR) where this emission is excited is $O(10^4)$yr. 

Finally, it is useful to consider the stark differences between objects embedded in AGN disks on prograde and on retrograde orbits. Objects on prograde orbits undergo Type I migration due to gas torques \citep[e.g.][]{McK12} leading to BH-BH, BH-NS, NS-NS mergers detectable with LIGO-Virgo \citep{McK14,Bartos17,Stone17,Tagawa20,McK20a}. Stars on prograde orbits can turn into undying behemoths, polluting the disk with metals, due to continuous accretion of fresh gas from the disk \citep{Leigh16b,MatteoAdam20,MelvynDoug20}. Objects on retrograde orbits in AGN experience a much weaker gas torque \citep{McK14}, but face a scouring headwind that allows stellar orbits to evolve rapidly (particularly early in the life of an AGN) as we show above. 

The final stages of orbital collapse for massive stars may be detectable in the less optically thick inner regions of the AGN disk. Since the orbits of retrograde stars quickly collapse into the SMBH, this suggests that the ionization of binaries due to three-body interactions \citep[e.g.][]{leigh18,leigh18b,barrera20} (and therefore the suppression of the compact object merger rate) by embedded objects on retrograde orbits is mostly a consideration in the early stages of AGN ($<0.1$\, Myr) \citep{Secunda20b}. 

\section{Conclusions}
As active galactic nuclei (AGN) `turn on', some stars will be embedded in accretion disks around SMBHs on retrograde orbits. Such stars experience strong headwinds,  drag, ablation and orbital evolution on short timescales, leading to a starfall event in the first O(0.1)Myr of an AGN. Long-lived bow-shock flares due to starfall slowly rise in luminosity to O($10^{42}{\rm erg/s}(R_{\ast}/R_{\odot})^{2}$) as angular momentum is lost and these flares may be detectable in optically thin inner disks for massive infalling stars. If the population of retrograde stars losing angular momentum interacts at small disk radii, binaries can form and scatterings can occur on short timescales. An AGN loss-cone is filled when  (2+1) scatterings are directed towards the central SMBH. Around SMBH $\leq 10^{8}M_{\odot}$, tidal disruption events (TDEs) can occur, either prograde or retrograde,  compared to inner disk gas.

The two classes of AGN TDEs have qualitatively different lightcurves. Prograde AGN TDEs are expected to have observed optical/UV TDE source-frame luminosities and decay times comparable to naked TDEs (peak $L_{\rm o/u} \sim {\rm O}(10^{43}){\rm erg/s}$ and decay over O(yr)). These TDEs should be followed by a higher AGN state post-TDE as the inner disk, including added ejecta mass, gradually loses angular momentum and accretes on the viscous timescale. Retrograde TDEs should have significantly higher luminosities than regular or prograde TDEs, with $L_{\rm o/u} \sim {\rm O}(10^{44}){\rm erg/s}$ or higher, with  decay O(months) to a lower AGN continuum state as the inner disk refills on the viscous timescale. Since these TDEs are the results of random scatterings, we expect approximately similar numbers of pro- and retro-grade TDEs in large samples of AGN.

{\section{Acknowledgements.}} BM acknowledges very useful discussions with Wenbin Lu and Brian Metzger on some of the ideas that form the basis of this paper. BM \& KESF are supported by NSF AST-1831415 and Simons Foundation Grant 533845.  N.W.C.L. gratefully acknowledges support from the Chilean government via Fondecyt Iniciacion Grant 11180005, and acknowledges financial support from Millenium Nucleus NCN19-058 (TITANs). The work of DS was carried out at the Jet Propulsion Laboratory, California Institute of Technology, under a contract with NASA. The Flatiron Institute is supported by the Simons Foundation. Thanks to Lucy Reading-Ikkanda for patiently assembling an excellent cartoon from our rough sketches.

\section*{Data Availability}
Any data used in this analysis are available on reasonable request from the first author (BM).

\bibliographystyle{mnras}
\bibliography{refs} % if your bibtex file is called example.bib

\end{document}